\documentclass[10pt]{article}
\usepackage{graphicx}
\usepackage{amsmath}
\usepackage{amssymb}
\usepackage{caption2}
\usepackage{multirow}
\begin{document}

\begin{center}
{\large {\bf \sc{Systematic analysis of the D-wave charmonium states with the QCD sum rules}}} \\[2mm]
Qi Xin \footnote{E-mail: xinqihd@163.com.  }, Zhi-Gang Wang \footnote{E-mail: zgwang@aliyun.com.  }

Department of Physics, North China Electric Power University, Baoding 071003, P. R. China\\
\end{center}

\begin{abstract}
We systematically study the 1D charmonium spin-triplet (with the $J^{PC}=1^{--}, 2^{--}, 3^{--}$) and spin-singlet (with the $J^{PC}=2^{-+}$)  via the QCD sum rules in comparison with the recent experimental results.  The predicted mass  $M_{\psi_1}=3.77\pm{0.09}\,\rm {GeV}$  supports identifying the $\psi_1$ as the $\psi(3770)$, the value
$M_{\psi_2}=3.82\pm{0.09}\,\rm {GeV}$  is consistent with the reported observation of the $\psi_2(3823)$, the prediction $M_{\psi_3}=3.84\pm{0.08}\,\rm {GeV}$ supports identifying the $\psi_3$ as the $\psi_3(3842)$. Additionally, we estimate the unobserved $\eta_{c2}$ state lies at $3.83\pm{0.09}\,\rm {GeV}$, and suggest detection prospects in the future. More experimental data  will help us to unravel the mass spectrum of the charmonium states near the open-charm thresholds.
\end{abstract}



\section{Introduction}\
Charmonium, a flavorless meson made of a charmed quark and its antiquark, provides an ideal system for  studying  QCD up to  the non-perturbative regime. Its study offers us an excellent  topic to investigate strong interactions. Therefore, systematically  exploring   the charmonium states can deepen our understanding of the dynamics involving  the heavy quarknioum systems. Since the observation of the $J/\psi$ in 1974 \cite{E598-jpsi-1974,SLAC-jpsi-1974}, experimental  and theoretical physicists have actively wrestling with the  charmonium states. Prior to the discovery of the  $X(3872)$ in 2003 \cite{Belle-3872-2003}, many  charmonium states, such as the $\eta_c$, $\eta_c^\prime$, $J/\psi$, $\psi^\prime$, $h_c$, $\chi_{c0}$, $\chi_{c1}$, $\chi_{c2}$,  $\psi(3770)$, $\psi(4040)$, $\psi(4415)$, etc, were reported by the experimental collaborations \cite{SLAC-etac-1980,SLAC-etac2S-1980,SLAC-chic0-1977,SLAC-chic1-1975,SLAC-chic2-1976,DASP-psi4040-4160-1978,SLAC-psi4415-1976}, which can be classified in the traditional quark model.
In fact, as early as the 1980s, the theoretical physicists began to focus  on those particles and examined the mass spectrum, decay modes and collision processes in detail  \cite{Novikov-1977,Godfrey-1985,Ebert-2003,Eichten-2003}.

The $1{}^3D_1$ charmonium candidate  was observed with the mass $3772\pm 6\mbox{ MeV}$  in 1977 \cite{3770-SLAC-1977}. In 2004, the Belle collaboration firstly  reported the $\psi(3770)$ in the process $B^+ \to \psi(3770) K^+$ with the decay modes $\psi(3770)  \to D^0 \bar{D}^0 / D^+D^-$ \cite{3770-Belle-2004}. On the other hand, the BESIII collaboration  recently observed the $\mathcal{R}(3780)$ with the mass $M_{\mathcal{R}(3780)}=3778.7\pm 0.5\pm0.3\mbox{ MeV}$ and  the decay width $\Gamma_{\mathcal{R}(3780)}=20.8\pm 0.8\pm1.7\mbox{ MeV}$, which can also be interpreted as the $1{}^3D_1$ charmonium state \cite{3770-BESIII-2024}.

The D-wave charmonium with the quantum numbers $J^{PC}=2^{--}$ has been explored experimentally several times, the Belle collaboration observed an evidence of a resonance with statistical significance of $3.8 \sigma$ in the $\chi_{c1}\gamma$ final state, whose mass is $3823.1\pm 1.8\pm0.7\mbox{ MeV}$ \cite{Belle-3823-2013}, and the BESIII collaboration reported the  $X(3823)$ in the process $e^+e^- \to \pi^+\pi^- X(3823)\to \pi^+\pi^- \gamma \chi _{c1}$ with a statistical significance of $6.2\sigma$, the measured mass  is $3821.7 \pm 1.3 \pm 0.7\mbox{ MeV}$ \cite{BESIII-3823-2015}. In 2020, the LHCb collaboration measured the energy  gaps among the $\psi_2(3823)$, $\chi_{c1}(3872)$ and $\psi(2S)$ states \cite{LHCb-3823-2020}. While the BESIII collaboration precisely measured the mass of the $\psi_2(3823)$ to be $3823.12\pm0.43\pm0.13 \mbox{ MeV}/3824.5\pm2.4\pm1.0 \mbox{ MeV}$ \cite{BESIII-3823-2022-1,BESIII-3823-2022-2}.

In the decay modes $X(3842)\to D^0\bar{D}^0$ and $ D^+D^-$, a new narrow charmonium state named as $X(3842)$ was  detected by the LHCb collaboration with very high statistical significance. Its determined mass and width are $ 3842.71\pm0.16 \pm0.12 \mbox{ MeV}$ and $ 2.79 \pm0.51 \pm0.35 \mbox{ MeV}$, respectively \cite{3842-LHCb-2019}, which  imply that the $X(3842)$ state might be  the $\psi_3(1{}^3D_3)$ charmonium state with the spin-parity-charge-conjugation $J^{PC} = 3^{--}$. While the $1{}^1D_2$ charmonium state has not been observed yet.

In 2023, the BESIII collaboration observed three resonance  structures $\mathcal{R}(3760)$, $\mathcal{R}(3780)$ and $\mathcal{R}(3810)$ in the cross sections for the process $e^+e^- \to \rm{non\,\,open\,\, charm \,\,hadrons}$ with significances of $9.4\sigma$, $15.7\sigma$, and $9.8\sigma$, respectively \cite{BESIII-2023-3810}, they observed the $\mathcal{R}(3810)$  for the first time \cite{BESIII-2023-3810}, and confirmed the old particles $\mathcal{R}(3760)$ and $\mathcal{R}(3780)$ \cite{BESIII-2020-3760-3780}. The experimental masses and widths were measured to be $M_{\mathcal{R}(3760)} = 3761.7\pm2.2 \pm1.2 \mbox{ MeV}$,  $\Gamma_{\mathcal{R}(3760)} = 6.7 \pm11.1 \pm1.1 \mbox{ MeV}$, $M_{\mathcal{R}(3780)} = 3784.7\pm5.7 \pm1.6 \mbox{ MeV} $, $\Gamma_{\mathcal{R}(3780)} = 31.6 \pm11.9 \pm3.2 \mbox{ MeV}  $, $ M_{\mathcal{R}(3810)} = 3805.8\pm1.1 \pm2.7 \,(3805.8\pm1.1 \pm2.7)\mbox{ MeV} $, $\Gamma_{\mathcal{R}(3810)} = 11.6 \pm2.9 \pm1.9 \,(11.5 \pm2.8 \pm1.9) \mbox{ MeV} $ \cite{BESIII-2023-3810}. And those parameters were improved recently, $M_{\mathcal{R}(3760)} = 3751.9\pm3.8 \pm2.8 \mbox{ MeV}$,  $\Gamma_{\mathcal{R}(3760)} = 32.8 \pm5.8 \pm8.7  \mbox{ MeV}$, $M_{\mathcal{R}(3780)} = 3778.7\pm0.5 \pm0.3 \mbox{ MeV} $, $\Gamma_{\mathcal{R}(3780)} = 20.3 \pm0.8 \pm1.7 \mbox{ MeV}  $, $ M_{\mathcal{R}(3810)} = 3804.5\pm0.9 \pm0.9 \mbox{ MeV} $, $\Gamma_{\mathcal{R}(3810)} = 5.4 \pm3.5 \pm3.2  \mbox{ MeV}$ \cite{3770-BESIII-2024}.

In our previous works, we have adopted the QCD sum rules approach as a  valuable theoretical tool to describe the masses, pole residues, form-factors, hadronic coupling constants and decay widths for  the charmed  mesons $D$, $D^*$, $D_s$, $D_s^*$, $D_{0}^{*}(2400)$, $D_{s0}^{*}(2317)$, $D_{1}(2430)$, $D_{s1}(2460)$, $D_{s1}^*(2860)$, $D_{s3}^*(2860)$ \cite{Ds2860-WZG-2860,D-WZG-2015,Dwave-cs-WZG-2016}, charmonium or charmonium-like states $X(3842)$, $X(3872)$, $h_c(4000)$, $\chi_{c1}(4010)$, and so on \cite{3842-Dwavecc-YGLWZG-2019,tetra-WZG-2024,tetramole-WZG-2021,tetra-WZG-2019,
WZG-review,WangZG-landau-PRD}. The masses of those charmonium-like states lie near the open-charm meson-pair thresholds, therefore they have been identified  as the charmonia, tetraquark states, molecular states etc, more theoretical and experimental investigations are still  needed to understand the mass spectrum of the charmonium states.

The S-wave and P-wave charmonium states  have been comprehensively explored in the framework of the QCD sum rules, we summarize the interpolating currents  in Table \ref{SPD-wave-charmonia}, however, the systematic study of the D-wave charmonium states via the QCD sum rules method is still scarce \cite{3842-Dwavecc-YGLWZG-2019}, we would  focus on this subject.

 For other theoretical works on the  D-wave charmonium  mass spectrum, one can consult the Bethe-Salpeter equations \cite{Fischer-charmonia-2015}, nonrelativistic potential model \cite{Barnes-D-wave-charmonia-2005} (include the screened potential model \cite{CKT-D-wave-charmonia-2009,ZXH-D-wave-charmonia-2017}, the linear potential model \cite{ZXH-D-wave-charmonia-2017,Bernardini-D-wave-charmonia-2003}, the Coulomb plus linear potential model \cite{Chaturvedi-D-wave-charmonia-2018}), the
relativized Godfrey-Isgur model \cite{Barnes-D-wave-charmonia-2004,Barnes-D-wave-charmonia-2005}, the relativistic screened potential model \cite{Bokade-D-wave-charmonia-2024}, the unquenched potential model \cite{LX-D-wave-charmonia-2019}, the Cornell potential (coupled-channel) model (with relativistic correction) \cite{Eichten-2003,Soni-D-wave-charmonia-2018}(\cite{Chaturvedi-D-wave-charmonia-2020}), the coupled-channel model \cite{LYR-D-wave-charmonia-2024}, the QCD-motivated relativistic quark model based on the quasipotential approach \cite{Ebert-D-wave-charmonia-2011} and the other potential model \cite{Sultan-D-wave-charmonia-2014,Kher-D-wave-charmonia-2018,
Bhavsar-D-wave-charmonia-2018,Radford-D-wave-charmonia-2007}.

\begin{table}
\begin{center}
\begin{tabular}{|c|c|c|c|c|c|c|c|c|}\hline\hline
\rm{L}                         & states         &$J^{PC}$      &currents        \\ \hline
\multirow{2}{*}{S-wave}   &$\eta_c$        &$0^{-+}$      &$\bar c(x)i\gamma_5 c(x)$          \\
                          &$J/\psi$        &$1^{--}$      &$\bar c(x)\gamma_\mu c(x)$  \\  \hline
\multirow{4}{*}{P-wave}   &$\chi_{c0}$     &$0^{++}$      &$\bar c(x)c(x)$         \\
                          &$\chi_{c1}$     &$1^{++}$      &$\bar c(x)\gamma_\mu \gamma_5c(x)$         \\
                          &$h_{c}$         &$1^{+-}$      &$\bar {c}(x)\sigma_{\mu\nu} c(x)$         \\
                          &$\chi_{c2}$     &$2^{++}$      &$\bar c(x)(\gamma_\mu \partial_\nu+\gamma_\nu \partial_\mu-\frac{1}{2}g_{\mu\nu}\!\not\!{\partial})c(x)$        \\ \hline
\multirow{4}{*}{D-wave}   &$\psi_1$        &$1^{--}$      &$J_{\mu}(x)$ in Eq.\eqref{currents}          \\
                          &$\psi_2$        &$2^{--}$      &$J^1_{\mu\nu}(x)$ in Eq.\eqref{currents}          \\
                          &$\eta_{c2}$     &$2^{-+}$      &$J^2_{\mu\nu}(x)$ in Eq.\eqref{currents}          \\
                          &$\psi_3$        &$3^{--}$       &$J_{\mu\nu\rho}(x)$ in Eq.\eqref{currents}   \\  \hline
\hline
\end{tabular}
\end{center}
\caption{The  quantum numbers and interpolating currents of the charmonia.}\label{SPD-wave-charmonia}
\end{table}

The  paper is arranged as follows: in Sect.2, we obtain  the QCD sum rules for the D-wave charmonium states; in Sect.3, we provide a comprehensive analysis; and in Sect.4, we give a short summary.

\section{QCD sum rules for the D-wave charmonium states}
The two-point correlation functions can be written as,
\begin{eqnarray}
\Pi_{\mu\nu}(p)&=&i\int d^4x e^{ip \cdot x} \langle
0|T\left\{J_\mu(x)J_\nu^{\dagger}(0)\right\}|0\rangle\, , \nonumber \\
\Pi_{\mu\nu\alpha\beta}(p)&=&i\int d^4x e^{ip \cdot x} \langle
0|T\left\{J_{\mu\nu}(x)J_{\alpha\beta}^{\dagger}(0)\right\}|0\rangle\, ,  \nonumber \\
\Pi_{\mu\nu\rho\alpha\beta\sigma}(p)&=&i\int d^4x e^{ip \cdot x} \langle
0|T\left\{J_{\mu\nu\rho}(x)J_{\alpha\beta\sigma}^{\dagger}(0)\right\}|0\rangle\, ,
\end{eqnarray}
where the interpolating currents, $J_{\mu\nu}(x)=J^1_{\mu\nu}(x)$, $J^2_{\mu\nu}(x)$,
\begin{eqnarray}\label{currents}
J_{\mu}(x)&=&\overline{c}(x)\stackrel{\leftrightarrow}{D}_\alpha \stackrel{\leftrightarrow}{D}_\beta\gamma_\rho \left(g^{\alpha\beta} g^{\rho\mu}+g^{\alpha\rho} g^{\beta\mu}+g^{\rho\beta} g^{\alpha\mu} \right) c(x) \, , \nonumber \\
J^1_{\mu\nu}(x)&=&\overline{c}(x)\left( \gamma_\mu\gamma\cdot\stackrel{\leftrightarrow}{D} \stackrel{\leftrightarrow}{D}_\nu +\gamma_\mu\stackrel{\leftrightarrow}{D}_\nu\gamma\cdot\stackrel{\leftrightarrow}{D} +\gamma_\nu\gamma\cdot\stackrel{\leftrightarrow}{D} \stackrel{\leftrightarrow}{D}_\mu +\gamma_\nu\stackrel{\leftrightarrow}{D}_\mu\gamma\cdot\stackrel{\leftrightarrow}{D} - g_{\mu\nu}\gamma\cdot\stackrel{\leftrightarrow}{D} \gamma\cdot\stackrel{\leftrightarrow}{D} \right)\gamma_5 c(x) \, , \nonumber \\
J^2_{\mu\nu}(x)&=&\overline{c}(x)\left( \stackrel{\leftrightarrow}{D}_\mu\stackrel{\leftrightarrow}{D}_\nu
+\stackrel{\leftrightarrow}{D}_\nu\stackrel{\leftrightarrow}{D}_\mu -\frac{1}{2} g_{\mu\nu}\stackrel{\leftrightarrow}{D}\cdot \stackrel{\leftrightarrow}{D} \right)\gamma_5 c(x) \, , \nonumber \\
J_{\mu\nu\rho}(x)&=&\overline{c}(x)\left( \stackrel{\leftrightarrow}{D}_\mu\stackrel{\leftrightarrow}{D}_\nu \gamma_\rho
+\stackrel{\leftrightarrow}{D}_\rho\stackrel{\leftrightarrow}{D}_\mu \gamma_\nu+\stackrel{\leftrightarrow}{D}_\nu\stackrel{\leftrightarrow}{D}_\rho \gamma_\mu\right) c(x) \, ,
\end{eqnarray}
 the covariant derivative $\stackrel{\leftrightarrow}{D}_\mu=\stackrel{\rightarrow}{\partial}_\mu-ig_sG_\mu-\stackrel{\leftarrow}{\partial}_\mu-ig_sG_\mu $, and the $G_\mu$ is the gluon field. We take
two-$\stackrel{\leftrightarrow}{D}_\mu$ to represent the D-wave to construct the currents $J_{\mu}(x)$, $J^1_{\mu\nu}(x)$, $J^2_{\mu\nu}(x)$ and $J_{\mu\nu\rho}(x)$ to interpolate the charmonium states $\psi_1$, $\psi_2$, $\eta_{c2}$ and $\psi_3$ respectively.  The quark currents with the covariant derivative $\stackrel{\leftrightarrow}{D}_\mu$ are gauge invariant, although which blurs the physical interpretation of the $\stackrel{\leftrightarrow}{D}_\mu$ being the angular momentum.

To obtain the hadronic representation, we insert a complete set of intermediate hadronic states with the same quantum numbers  as the current operators $J_{\mu}(x)$ $J_{\mu\nu}(x)$ and $J_{\mu\nu\rho}(x)$ into the correlation functions  $\Pi_{\mu\nu}(p)$, $\Pi_{\mu\nu\alpha\beta}(p)$ and $\Pi_{\mu\nu\rho\alpha\beta\sigma}(p)$ \cite{ Dwave-cs-WZG-2016,ZhuJJ-tensor,QCDSR-SVZ79,QCDSR-Reinders85}, and consider the current-hadron couplings,
 \begin{eqnarray}
\langle 0|J_{\mu }(0)|\psi_{1}(p)\rangle &=&f_{\psi_{1}}\varepsilon_{\mu}\, , \nonumber\\
 \langle 0| J_{\mu}(0)|\chi_{c0}(p)\rangle &=& f_{\chi_{c0}}  p_\mu \, ,
\end{eqnarray}
 \begin{eqnarray}
 \langle 0|J_{\mu\nu}(0)|\psi_{2}/\eta_{c2}(p)\rangle&=&f_{\psi_{2}/\eta_{c2}}\varepsilon_{\mu\nu} \, , \nonumber\\
\langle  0|J_{\mu\nu}(0)|\chi_{c1}/h_c(p)\rangle&=&
f_{\chi_{c1}/h_c}\left(p_\mu\varepsilon_{\nu}+p_\nu\varepsilon_{\mu} \right)\, , \nonumber\\
  \langle 0|J_{\mu\nu}(0)|\eta_c(p)\rangle&=&f_{\eta_c}p_{\mu} p_{\nu}\, ,
\end{eqnarray}
 \begin{eqnarray}
 \langle 0|J_{\mu\nu\rho}(0)|\psi_{3}(p)\rangle&=&f_{\psi_{3}}\varepsilon_{\mu\nu\rho} \, ,\nonumber \\
\langle 0| J_{\mu\nu\rho}(0)|\chi_{c2}(p)\rangle &=&f_{\chi_{c2}} ( p_\mu \varepsilon_{\nu\rho}+p_\nu\varepsilon_{\rho\mu}+p_\rho\varepsilon_{\mu\nu}) \, , \nonumber\\
\langle 0| J_{\mu\nu\rho}(0)|J/\psi/\psi_1(p)\rangle &=&f_{J/\psi/\psi_1} ( p_\mu p_\nu \varepsilon_{\rho}+p_\nu p_\rho \varepsilon_{\mu}+p_\rho p_\mu\varepsilon_{\nu}) \, , \nonumber\\
\langle 0| J_{\mu\nu\rho}(0)|\chi_{c0}(p)\rangle &=&f_{\chi_{c0}}  p_\mu p_\nu p_{\rho} \, ,
\end{eqnarray}
where the $f_{\psi_{1}}$, $f_{\psi_{2}}$, $f_{\eta_{c2}}$, $f_{\psi_{3}}$, $f_{\chi_{c0}}$, $f_{\chi_{c2}}$ and $f_{J/\psi}$ are  decay constants, the $\varepsilon_{\mu}$, $\varepsilon_{\mu\nu}$ and $\varepsilon_{\mu\nu\rho}$ are the polarization vectors of the charmonium states. Then we perform detailed  tensor analysis  and isolate the ground state contributions,
\begin{eqnarray}
\Pi_{\mu\nu}(p)&=&\Pi(p^2)\,\widetilde{g}_{\mu\nu}+\Pi_0(p^2)\, p_\mu p_\nu  \, , \nonumber\\
&&=\frac{f_{\psi_{1}}^2}{M_{\psi_{1}}^2-p^2}\,\widetilde{g}_{\mu\nu}+ ...\, ,\nonumber\\
  \Pi_{\mu\nu\alpha\beta}(p)&=&\Pi(p^2){\rm P}_{\mu\nu\alpha\beta} +\Pi_1(p^2)\left( \widetilde{g}_{\mu \alpha}p_{ \nu}  p_{ \beta}+\widetilde{g}_{\mu \beta}p_{ \nu}  p_{ \alpha}+ \widetilde{g}_{\nu \alpha}p_{ \mu}  p_{ \beta}+ \widetilde{g}_{\nu \beta}p_{ \mu}  p_{ \alpha} \right)+\Pi_0(p^2)\, p_{\mu }p_{ \nu} p_{\alpha } p_{ \beta}  \, ,\nonumber\\
&&=\frac{f_{\psi_{2}/\eta_{c2}}^2}{M_{\psi_{2}/\eta_{c2}}^2-p^2}{\rm P}_{\mu\nu\alpha\beta}+ ...\, ,\nonumber\\
  \Pi_{\mu\nu\rho\alpha\beta\sigma}(p)&=&\Pi(p^2){\rm P}_{\mu\nu\rho\alpha\beta\sigma} +\Pi_2(p^2)\left( {\rm P}_{\nu\rho\beta\sigma}\,p_\mu p_\alpha + {\rm P}_{\nu\rho\alpha\sigma}\,p_\mu p_\beta + {\rm P}_{\nu\rho\alpha\beta}\,p_\mu p_\sigma + {\rm P}_{\mu\rho\beta\sigma}\,p_\nu p_\alpha\right.\nonumber\\
&&\left.+ {\rm P}_{\mu\rho\alpha\sigma}\,p_\nu p_\beta+ {\rm P}_{\mu\rho\alpha\beta}\,p_\nu p_\sigma+ {\rm P}_{\mu\nu\beta\sigma}\,p_\rho p_\alpha+ {\rm P}_{\mu\nu\alpha\sigma}\,p_\rho p_\beta+ {\rm P}_{\mu\nu\alpha\beta}\,p_\rho p_\sigma\right) \nonumber\\
&&+\Pi_1(p^2)\left(\widetilde{g}_{\mu \alpha}\, p_\nu p_\rho p_\beta p_\sigma+\widetilde{g}_{\mu \beta}\, p_\nu p_\rho p_\alpha p_\sigma
+\widetilde{g}_{\mu \sigma}\, p_\nu p_\rho p_\alpha p_\beta +\widetilde{g}_{\nu \alpha}\, p_\mu p_\rho p_\beta p_\sigma\right.\nonumber\\
&&\left.+\widetilde{g}_{\nu \beta} \,p_\mu p_\rho p_\alpha p_\sigma+\widetilde{g}_{\nu \sigma}\, p_\mu p_\rho p_\alpha p_\beta
+\widetilde{g}_{\rho \alpha} \, p_\mu p_\nu p_\beta p_\sigma+\widetilde{g}_{\rho \beta}\, p_\mu p_\nu p_\alpha p_\sigma
+\widetilde{g}_{\rho \sigma}\, p_\mu p_\nu p_\alpha p_\beta\right) \nonumber\\
&&+\Pi_0(p^2)\, p_\mu p_\nu p_\rho p_\alpha p_\beta p_\sigma \, , \nonumber\\
&&=\frac{f_{\psi_{3}}^2}{M^2_{\psi_{3}}-p^2}{\rm P}_{\mu\nu\rho\alpha\beta\sigma}+ ...\, ,
\end{eqnarray}
where
\begin{eqnarray}
\widetilde{g}_{\mu\nu} &=&-g_{\mu\nu}+\frac{p_\mu p_\nu}{p^2}  \, , \nonumber\\
  {\rm P}_{\mu\nu\alpha\beta}&=& \frac{\widetilde{g}_{\mu\alpha}\widetilde{g}_{\nu\beta}
 +\widetilde{g}_{\mu\beta}\widetilde{g}_{\nu\alpha}}{2}-\frac{\widetilde{g}_{\mu\nu}\widetilde{g}_{\alpha\beta}}{3}  \, , \nonumber\\
 {\rm P}_{\mu\nu\rho\alpha\beta\sigma}&=&\frac{1}{6}\left(\widetilde{g}_{\mu\alpha}\widetilde{g}_{\nu\beta}\widetilde{g}_{\rho\sigma}+\widetilde{g}_{\mu\alpha}\widetilde{g}_{\nu\sigma}\widetilde{g}_{\rho\beta}
+\widetilde{g}_{\mu\beta}\widetilde{g}_{\nu\alpha}\widetilde{g}_{\rho\sigma}   +\widetilde{g}_{\mu\beta}\widetilde{g}_{\nu\sigma}\widetilde{g}_{\rho\alpha}
 +\widetilde{g}_{\mu\sigma}\widetilde{g}_{\nu\alpha}\widetilde{g}_{\rho\beta}+\widetilde{g}_{\mu\sigma}\widetilde{g}_{\nu\beta}\widetilde{g}_{\rho\alpha}\right)\nonumber\\
&&-\frac{1}{15}\left(\widetilde{g}_{\mu\alpha}\widetilde{g}_{\nu\rho}\widetilde{g}_{\beta\sigma}+\widetilde{g}_{\mu\beta}\widetilde{g}_{\nu\rho}\widetilde{g}_{\alpha\sigma}
+\widetilde{g}_{\mu\sigma}\widetilde{g}_{\nu\rho}\widetilde{g}_{\alpha\beta}  +\widetilde{g}_{\nu\alpha}\widetilde{g}_{\mu\rho}\widetilde{g}_{\beta\sigma}
 +\widetilde{g}_{\nu\beta}\widetilde{g}_{\mu\rho}\widetilde{g}_{\alpha\sigma}   +\widetilde{g}_{\nu\sigma}\widetilde{g}_{\mu\rho}\widetilde{g}_{\alpha\beta}\right. \nonumber\\
&&\left. +\widetilde{g}_{\rho\alpha}\widetilde{g}_{\mu\nu}\widetilde{g}_{\beta\sigma}  +\widetilde{g}_{\rho\beta}\widetilde{g}_{\mu\nu}\widetilde{g}_{\alpha\sigma}
         +\widetilde{g}_{\rho\sigma}\widetilde{g}_{\mu\nu}\widetilde{g}_{\alpha\beta}\right) \, .
\end{eqnarray}
We choose the components $\Pi(p^2)$ to analyze the charmonium states with the quantum numbers $J^{PC}=1^{--}$, $2^{--}$, $2^{-+}$ and $3^{--}$ without contaminations.

\begin{figure}
 \centering
   \includegraphics[totalheight=2cm,width=2.5cm]{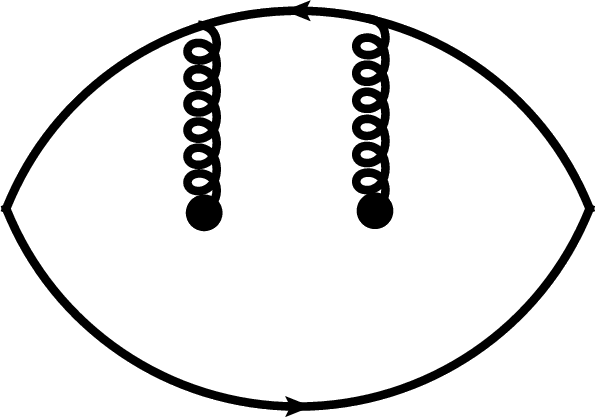}
    \includegraphics[totalheight=2cm,width=2.5cm]{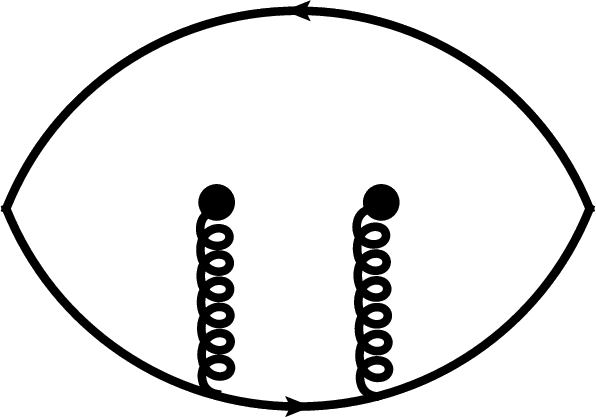}
    \includegraphics[totalheight=2cm,width=2.5cm]{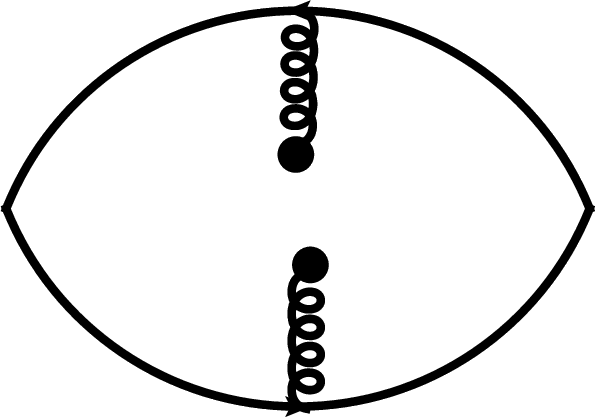}
    \includegraphics[totalheight=2cm,width=2.5cm]{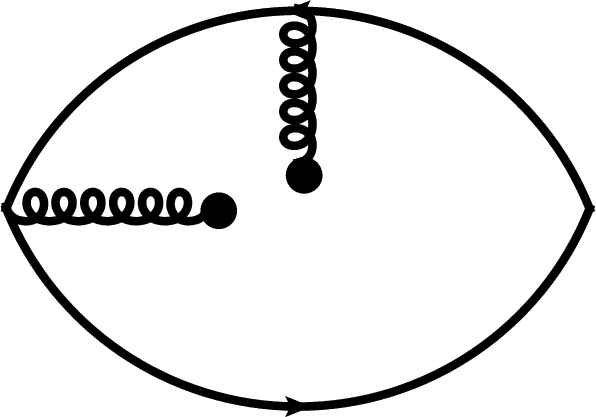}
    \includegraphics[totalheight=2cm,width=2.5cm]{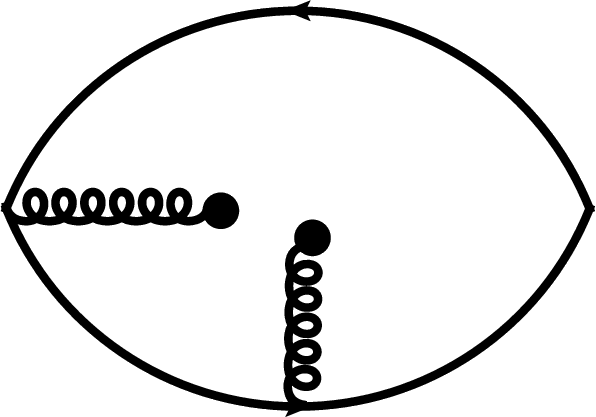}
      \includegraphics[totalheight=2cm,width=2.5cm]{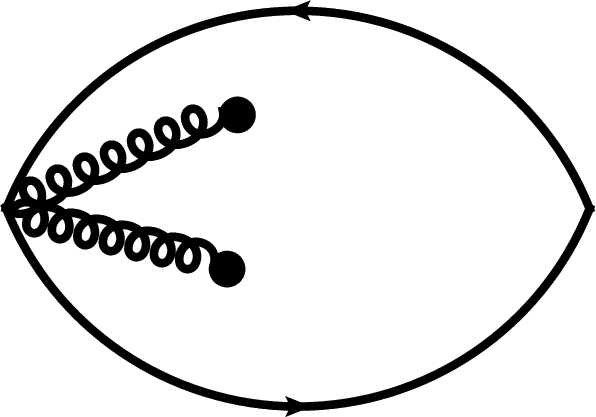}
    \caption{The Feynman diagrams contribute to the gluon condensate $\langle \frac{\alpha_sGG}{\pi}\rangle$. }\label{gluon-condensate}
\end{figure}

\begin{figure}
 \centering
   \includegraphics[totalheight=2cm,width=2.5cm]{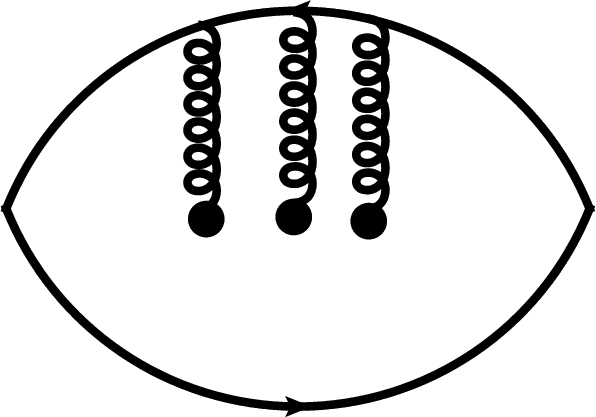}
    \includegraphics[totalheight=2cm,width=2.5cm]{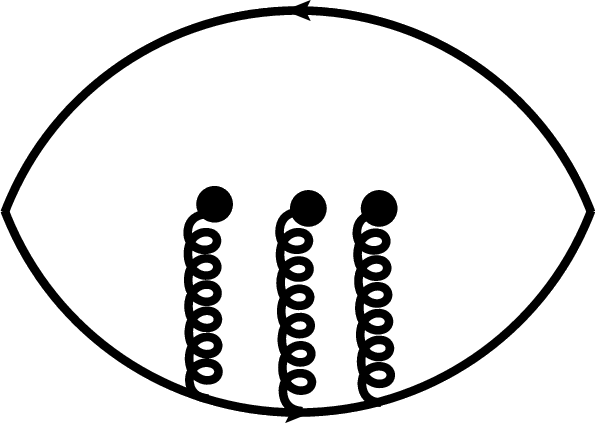}
    \includegraphics[totalheight=2cm,width=2.5cm]{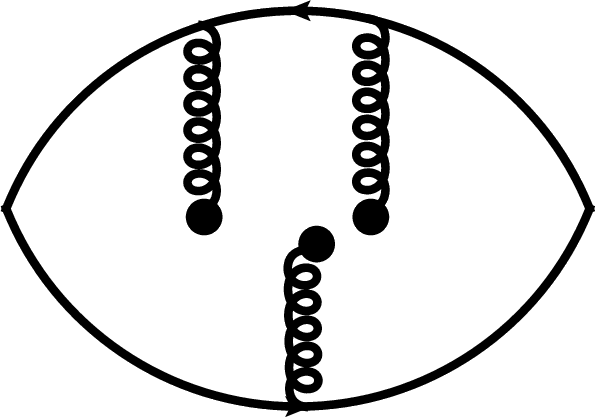}
    \includegraphics[totalheight=2cm,width=2.5cm]{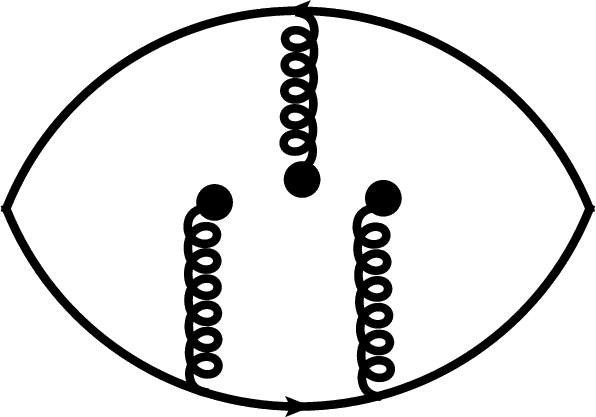}
    \includegraphics[totalheight=2cm,width=2.5cm]{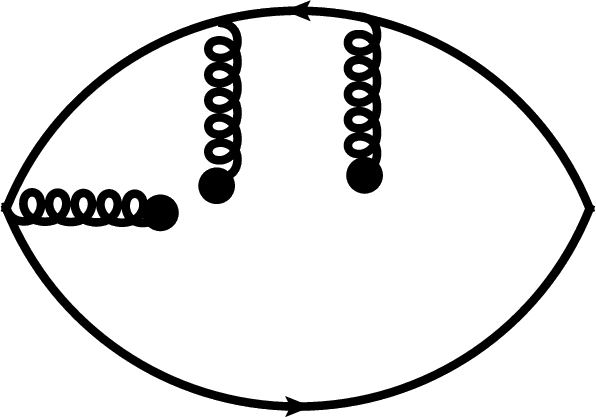}
    \includegraphics[totalheight=2cm,width=2.5cm]{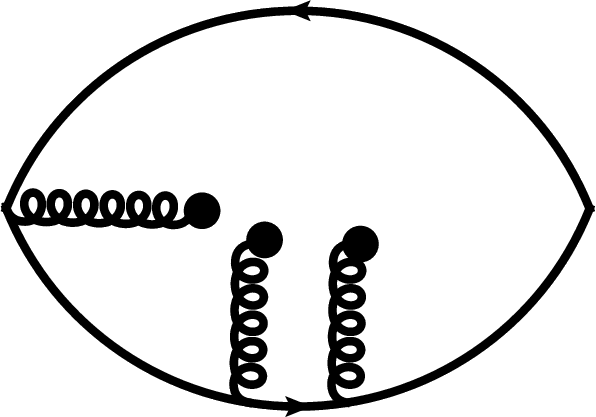}
    \includegraphics[totalheight=2cm,width=2.5cm]{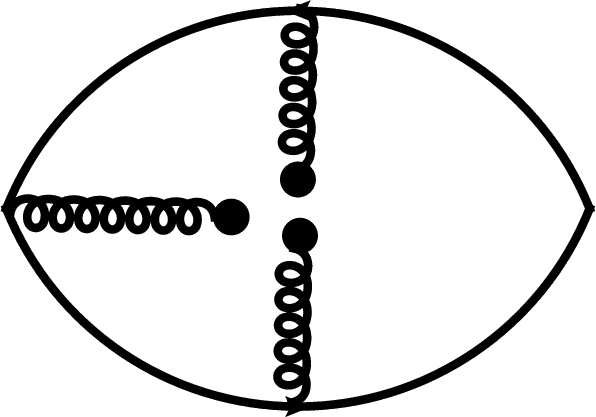}
    \includegraphics[totalheight=2cm,width=2.5cm]{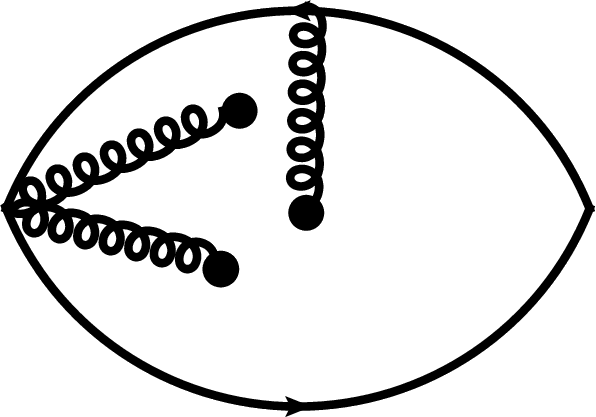}
    \includegraphics[totalheight=2cm,width=2.5cm]{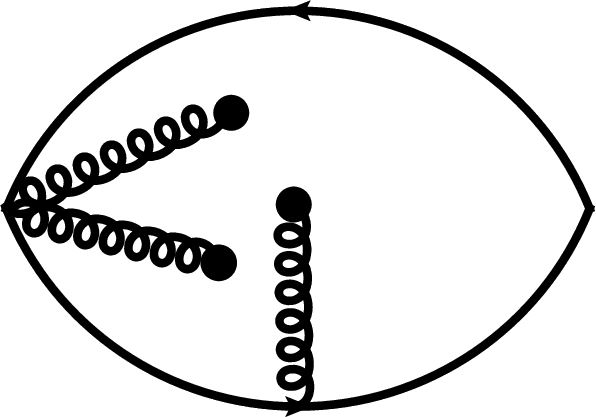}
    \caption{The Feynman diagrams contribute to the three-gluon condensate $\langle g_s^3 GGG\rangle$. }\label{three-gluon-condensate}
\end{figure}

At the QCD side, we accomplish  the operator product expansion up to the gluon condensate $\langle \frac{\alpha_s GG}{\pi}\rangle$ and three-gluon condensate $\langle g_s^3GGG\rangle$, and the corresponding Feynman diagrams  are displayed in Figs.\ref{gluon-condensate}-\ref{three-gluon-condensate}. In computation, we evaluate the integrals in momentum space, select the components $\Pi(p^2)$ associated with the particular structures in the correlation functions  $\Pi_{\mu\nu}(p)$, $\Pi_{\mu\nu\alpha\beta}(p)$ and $\Pi_{\mu\nu\rho\alpha\beta\sigma}(p)$  to investigate the charmonium states with the spin-parity $J^P=1^-$, $2^-$ and $3^-$ respectively. At last, we  determine  the QCD spectral density $\rho(s)$  using dispersion relation.

We take quark-hadron duality below the continuum threshold $s_0$ and perform the Borel transformation  with respect to the variable $P^2=-p^2$ to acquire  four QCD sum rules,
\begin{eqnarray}\label{QCDSR-M}
f^2\exp\left( -\frac{M^2}{T^2}\right) &=&\int_{4m_c^2}^{s_0}ds \,\rho(s)\exp\left( -\frac{s}{T^2}\right)  \, ,
\end{eqnarray}
where the $T^2$ are the Borel parameters, the decay constants $f=f_{\psi_1}$, $f_{\psi_2}$, $f_{\eta_{c2}}$ and $f_{\psi_3}$,  the QCD spectral densities $\rho(s)=\rho_1(s)$, $\rho_2^1(s)$, $\rho_2^2(s)$ and $\rho_3(s)$,
\begin{eqnarray}\label{Density-1}
\rho_1(s)&=&\frac{3(s-4m_c^2)^2\sqrt{s(s-4m_c^2)}}{8\pi^2}+\langle \frac{\alpha_s GG}{\pi}\rangle\frac{24m_c^6-51sm_c^4+35s^2m_c^2-5s^3}{6s\sqrt{s(s-4m_c^2)}} \nonumber\\
&&+\langle g_s^3GGG\rangle\frac{256m_c^6-1160sm_c^4+566s^2m_c^2-83s^3}{192\pi^2s(s-4m_c^2)\sqrt{s(s-4m_c^2)}}\, ,
 \end{eqnarray}

\begin{eqnarray}
\rho_2^1(s)&=&\frac{4(s-4m_c^2)^3(s+6m_c^2)}{5\pi^2\sqrt{s(s-4m_c^2)}}+\langle \frac{\alpha_s GG}{\pi}\rangle\frac{16(38m_c^6-61sm_c^4-s^2m_c^2+3s^3)}{9s\sqrt{s(s-4m_c^2)}} \nonumber\\
&&+\langle g_s^3GGG\rangle\frac{-192m_c^8+928sm_c^6-372s^2m_c^4+51s^3m_c^2-4s^4}{18\pi^2s^2(s-4m_c^2)\sqrt{s(s-4m_c^2)}}\, ,
 \end{eqnarray}

\begin{eqnarray}
\rho_2^2(s)&=&\frac{(s-4m_c^2)^2\sqrt{s(s-4m_c^2)}}{5\pi^2}+\langle \frac{\alpha_s GG}{\pi}\rangle\frac{2(16m_c^6-30sm_c^4-3s^2m_c^2+2s^3)}{9s\sqrt{s(s-4m_c^2)}} \nonumber\\
&&+\langle g_s^3GGG\rangle\frac{2m_c^4+3sm_c^2}{6\pi^2s\sqrt{s(s-4m_c^2)}}\, ,
\end{eqnarray}

\begin{eqnarray}\label{Density-4}
\rho_3(s)&=&\frac{9(s-4m_c^2)^3(s+3m_c^2)}{35\pi^2\sqrt{s(s-4m_c^2)}}+\langle \frac{\alpha_s GG}{\pi}\rangle\frac{120m_c^6-92sm_c^4+34s^2m_c^2-5s^3}{2s\sqrt{s(s-4m_c^2)}} \nonumber\\
&&+\langle g_s^3GGG\rangle\frac{576m_c^8-992sm_c^6+284s^2m_c^4+6s^3m_c^2-7s^4}{16\pi^2s^2(s-4m_c^2)\sqrt{s(s-4m_c^2)}}\, ,
 \end{eqnarray}
 the $\rho_1(s)$, $\rho_2^1(s)$, $\rho_2^2(s)$ and $\rho_3(s)$ correspond the quark  currents  $J_{\mu}(x)$, $J_{\mu\nu}^1(x)$, $J_{\mu\nu}^2(x)$, and $J_{\mu\nu\rho}(x)$, respectively.

We extract the QCD sum rules for the masses  of the D-wave charmonium states $\psi_1$, $\psi_2$, $\eta_{c2}$ and $\psi_3$ by differentiating Eq.\eqref{QCDSR-M} with respect to $\frac{1}{T^2}$ and eliminating the decay constants,
 \begin{eqnarray}
 M^2 &=& -\frac{\frac{d }{d\tau} \int_{4m_c^2}^{s_0}ds \,\rho(s)\exp\left(- \tau s\right)}{\int_{4m_c^2}^{s_0}ds \,\rho(s)\exp\left(- \tau s\right)} \mid_{\tau=\frac{1}{T^2}}\, .
 \end{eqnarray}

\section{Numerical results and discussions}
We  adopt  the $\overline{MS}$
(modified-minimal-subtraction) mass   $m_{c}(m_c)=(1.275\pm0.025)\,\rm{GeV}$
 (from the Particle Data Group), which evolves with the energy scale $\mu$ according to the re-normalization  group equation \cite{PDG},
\begin{eqnarray}
m_c(\mu)&=&m_c(m_c)\left[\frac{\alpha_{s}(\mu)}{\alpha_{s}(m_c)}\right]^{\frac{12}{33-2n_f}} \, ,\nonumber\\
\alpha_s(\mu)&=&\frac{1}{b_0t}\left[1-\frac{b_1}{b_0^2}\frac{\log t}{t} +\frac{b_1^2(\log^2{t}-\log{t}-1)+b_0b_2}{b_0^4t^2}\right]\, ,
\end{eqnarray}
where $t=\log \frac{\mu^2}{\Lambda_{QCD}^2}$, $b_0=\frac{33-2n_f}{12\pi}$, $b_1=\frac{153-19n_f}{24\pi^2}$, $b_2=\frac{2857-\frac{5033}{9}n_f+\frac{325}{27}n_f^2}{128\pi^3}$,  $\Lambda_{QCD}=210\,\rm{MeV}$, $292\,\rm{MeV}$  and  $332\,\rm{MeV}$ for the flavors  $n_f=5$, $4$ and $3$, respectively  \cite{PDG},  and we take the flavor $n_f=4$ and
 $m_c=\mu=(1.275\pm0.025)\,\rm{GeV}$.  We adopt two sets of parameters for the gluon condensate  $\langle \frac{\alpha_s GG}{\pi}\rangle$ and three-gluon condensate $\langle g_s^3GGG\rangle$, as listed in Table \ref{parameter}.

\begin{table}
\begin{center}
\begin{tabular}{|c|c|c|} \hline\hline
Input parameters & (I)\, \cite{QCDSR-SVZ79,QCDSR-Reinders85,QCDSR-Colangelo-Review}               & (II)\, \cite{QCDSR-Narison-I,QCDSR-Narison-II}    \\ \hline
$\langle\frac{\alpha_s GG}{\pi}\rangle \,(\rm{GeV}^4) $      &$0.012$\,
& $0.022\pm0.004 $ \,  \\ \hline
$\langle g_s^3GGG\rangle \,(\rm{GeV}^6)$                    & $0.045$               & $0.616\pm0.385$ \\ \hline\hline
\end{tabular}
\caption{The QCD input parameters, where the I and II denote the different values of the gluon condensate and three-gluon condensate.}\label{parameter}
\end{center}
\end{table}

The pole contributions are defined as,
\begin{eqnarray}
\rm{ pole} &=&\frac{\int_{4m_c^2}^{s_0}ds \,\rho\left(s\right) \exp\left(-\frac{s}{T^2}\right)}{\int_{4m_c^2}^{\infty}ds \,\rho\left(s\right) \exp\left(-\frac{s}{T^2}\right)} \, .
\end{eqnarray}
In order to avoid parameter dependence to the utmost extent, we prefer the central value of the pole contributions $55\%$ for all the  charmonium states $\psi_1$, $\psi_2$, $\eta_{c2}$ and $\psi_3$. To avoid contaminations from the higher resonances and continuum states, we adopt  the constraint  $\sqrt{s_0} = M + 0.45\sim0.65\,\rm{GeV}$ for the continuum threshold parameters, and try to adjust the continuum threshold parameters to reproduce the charmonium masses $M$ via trial and error, as the largest energy gap between the ground state and first radial excited state is $0.65\,\rm{GeV}$ for the charmonium states  (i.e. $M_{\eta_c^\prime}-M_{\eta_c}=0.65\,\rm{GeV}$) \cite{PDG}. The selection of the continuous threshold parameters both refer to the experimental results, and require ensuring that the operator product expansion at QCD side, the hadron sides and QCD sides are matched through dispersion relation to get the properties of the hadron states. The primary criteria are internal stability of the QCD sum rules, pole dominance, convergence of the OPE, and the existence of the Borel platform. These criteria take precedence over importing precise energy gaps predicted by other models, because the QCD sum rules extract hadron properties directly from correlation functions without assuming a specific dynamical model.

\begin{figure}
 \centering
  \includegraphics[totalheight=5cm,width=7cm]{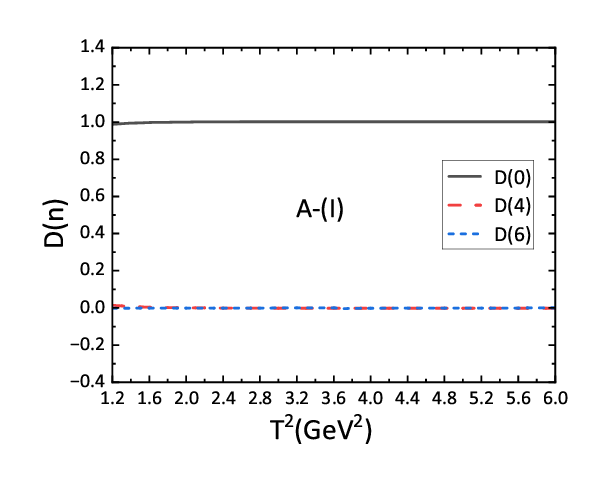}
   \includegraphics[totalheight=5cm,width=7cm]{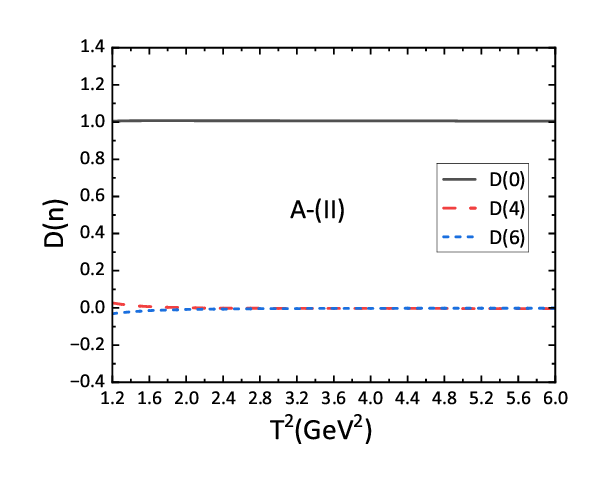}
   \includegraphics[totalheight=5cm,width=7cm]{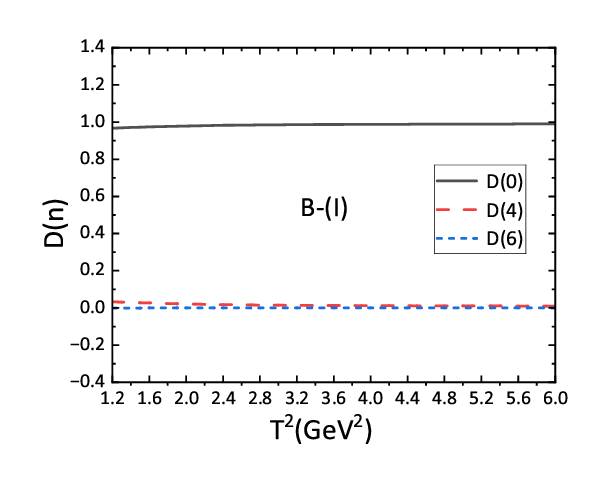}
   \includegraphics[totalheight=5cm,width=7cm]{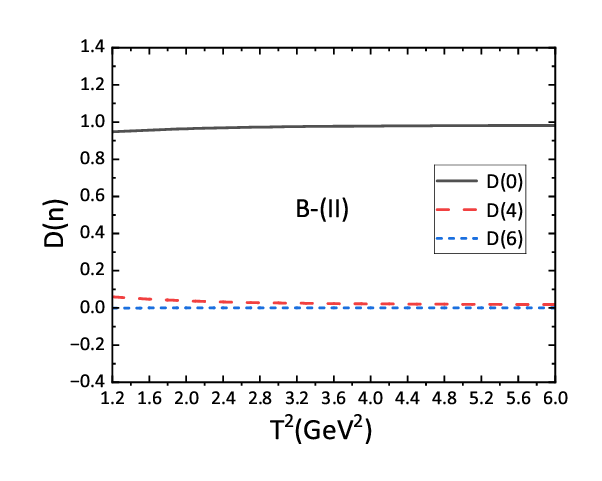}
   \includegraphics[totalheight=5cm,width=7cm]{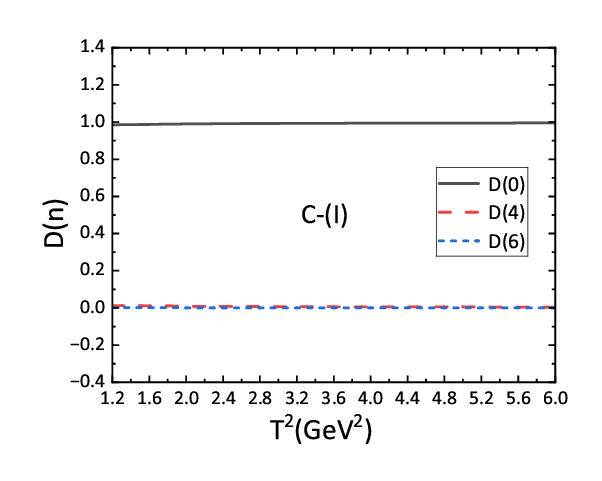}
   \includegraphics[totalheight=5cm,width=7cm]{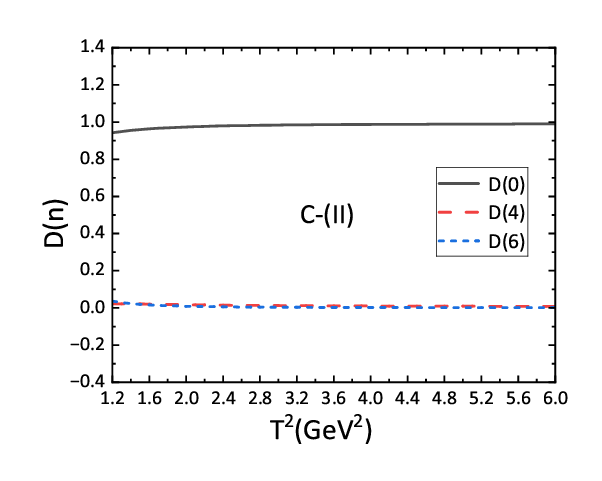}
     \includegraphics[totalheight=5cm,width=7cm]{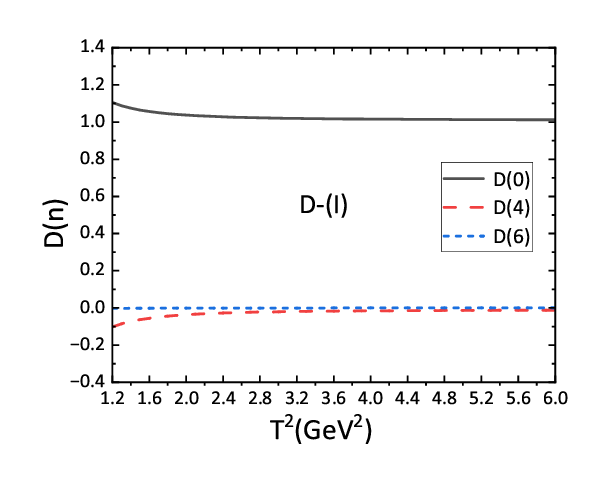}
     \includegraphics[totalheight=5cm,width=7cm]{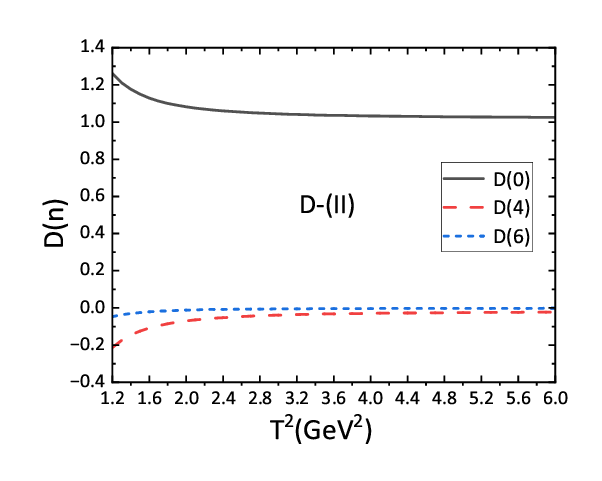}
    \caption{The contributions of the vacuum condensates of dimension $n$  with variations of the Borel parameters $T^2$ of the input parameters are set to central values, where the $A$, $B$, $C$ and $D$ denote $\psi_1$, $\psi_2$, $\eta_{c2}$ and $\psi_3$ charmonium states, respectively. }\label{Dn}
\end{figure}

In the QCD spectral densities given in Eqs.\eqref{Density-1}-\eqref{Density-4}, when the threshold $s=4m_c^2$ is reached, the endpoint singularity  arise from the factor the QCD spectral densities appear in the gluon condensate $\frac{1}{\sqrt{s(s-4m_c^2)}}$. Due to the minor contribution from the gluon condensate terms compared to the perturbative term as shown in Fig.\ref{Dn}, we attempt to implement the routine replacements $s=4m_c^2 \to s=4m_c^2+ \Delta^2$ with $\Delta ^2= \frac{m_c^2}{2}$, $m_c^2$, $\frac{3m_c^2}{2}$ in numerical computations. The determination of numerical results for masses at three different $\Delta^2$ values are consistent. To ensure as much consistency in parameters selection as possible, we adopted the $\Delta ^2= m_c^2$ substitution as used in our previous work \cite{WangZG-deta-2022,WangZG-deta-2021}, then the corresponding Borel parameters and masses are obtained.

We  define the normalized  contributions $D(n)$,
\begin{eqnarray}
D(n)&=&\frac{\int_{4m_{c}^{2}}^{s_{0}}ds\,\rho_{QCD,n}(s)\exp\left(-\frac{s}{T^{2}}\right)}
{\int_{4m_{c}^{2}}^{s_{0}}ds\,\rho_{QCD}\left(s\right)\exp\left(-\frac{s}{T^{2}}\right)}\, ,
\end{eqnarray}
where the $\rho_{QCD,n}(s)$ are the QCD spectral densities $\rho_1(s)$, $\rho_2^1(s)$, $\rho_2^2(s)$ and $\rho_3(s)$ involving the vacuum condensates of dimension $n$. In Fig.\ref{Dn}, we  plot  the contributions of the vacuum condensates $D(n)$ for the charmonium states $\psi_1$, $\psi_2$, $\eta_{c2}$ and $\psi_3$ for the central values of the input parameters given in Table \ref{parameter}, which reveals that the perturbative terms account for the major contributions, the three-gluon condensate $\langle g_s^3 GGG\rangle$ contributions are minimal, and they have the hierarchy   $|D(0)|\gg|D(4)|\geq|D(6)|$, the operator product expansion converges rather well.

\begin{table}
\begin{center}
\begin{tabular}{|c|c|c|c|c|c|c|c|c|}\hline\hline
$\psi$     &$J^{PC}$       & $T^2 (\rm{GeV}^2)$   &$\sqrt{s_0}(\rm {GeV}) $    &pole &$M (\rm{GeV})$     & $f (\rm{GeV}^4) $

\\ \hline

$\psi_1(1{}^3D_1)$          &$1^{--}$                       &$2.9-3.5$            &$4.30\pm0.1$              &$(43-68)\%$          &$3.77_{-0.09}^{+0.09}$      &$13.83_{-1.99}^{+2.17}$   \\  \hline

$\psi_2(1{}^3D_2)$        &$2^{--}$                        &$3.2-3.8$          &$4.40\pm0.1$              &$(48-70)\%$           &$3.82_{-0.09}^{+0.08}$            &$29.23_{-3.69}^{+4.00}$           \\  \hline

$\eta_{c2}(1{}^1D_2)$        &$2^{-+}$                  &$3.0-3.6$             &$4.40\pm0.1$            &$(47-71)\%$                &$3.83_{-0.08}^{+0.09}$        &$11.41_{-1.57}^{+1.72}$  \\   \hline

$\psi_3(1{}^3D_3)$         &$3^{--}$                    &$3.2-3.8$               &$4.40\pm0.1$              &$(45-68)\%$        &$3.84_{-0.08}^{+0.08}$      &$14.77_{-1.92}^{+2.09}$    \\

\hline\hline
\end{tabular}
\end{center}
\caption{The spin-parity-charge-conjugation, Borel parameters, continuum threshold parameters, pole contributions, masses, decay constants of the charmonium states for the input parameters I.}\label{mass-I}
\end{table}

\begin{table}
\begin{center}
\begin{tabular}{|c|c|c|c|c|c|c|c|c|}\hline\hline
$\psi$     &$J^{PC}$       & $T^2 (\rm{GeV}^2)$   &$\sqrt{s_0}(\rm {GeV}) $    &pole &$M (\rm{GeV})$     & $f (\rm{GeV}^4) $

\\ \hline

$\psi_1(1{}^3D_1)$          &$1^{--}$                       &$2.9-3.5$            &$4.30\pm0.1$              &$(43-68)\%$          &$3.77_{-0.08}^{+0.09}$      &$13.82_{-1.98}^{+2.18}$   \\  \hline

$\psi_2(1{}^3D_2)$        &$2^{--}$                        &$3.2-3.8$          &$4.40\pm0.1$              &$(48-70)\%$             &$3.82_{-0.09}^{+0.08}$            &$29.29_{-3.69}^{+4.00}$           \\  \hline

$\eta_{c2}(1{}^1D_2)$        &$2^{-+}$                  &$3.0-3.6$             &$4.40\pm0.1$            &$(47-71)\%$                &$3.83_{-0.09}^{+0.09}$        &$11.64_{-1.58}^{+1.72}$  \\   \hline

$\psi_3(1{}^3D_3)$         &$3^{--}$                 &$3.2-3.8$            &$4.40\pm0.1$              &$(45-68)\%$        &$3.84_{-0.07}^{+0.09}$      &$14.94_{-1.94}^{+2.11}$    \\

\hline\hline
\end{tabular}
\end{center}
\caption{The spin-parity-charge-conjugation, Borel parameters, continuum threshold parameters, pole contributions, masses, decay constants of the charmonium states for the input parameters II.}\label{mass-II}
\end{table}

\begin{figure}
 \centering
 \includegraphics[totalheight=5cm,width=7cm]{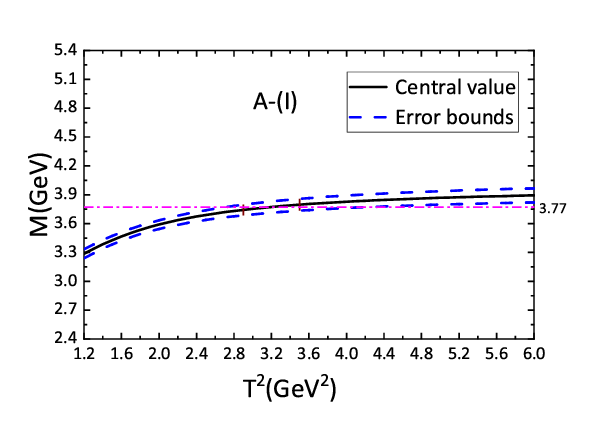}
  \includegraphics[totalheight=5cm,width=7cm]{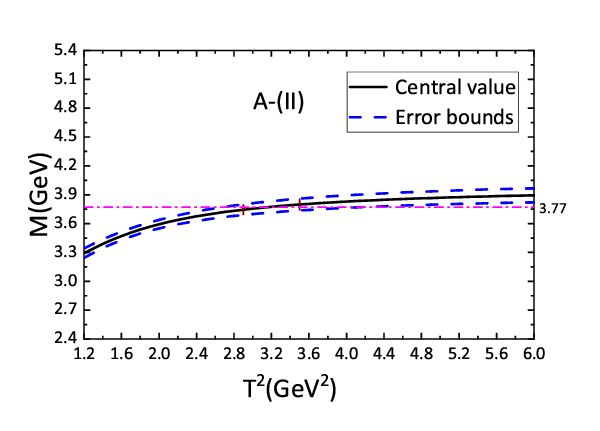}
   \includegraphics[totalheight=5cm,width=7cm]{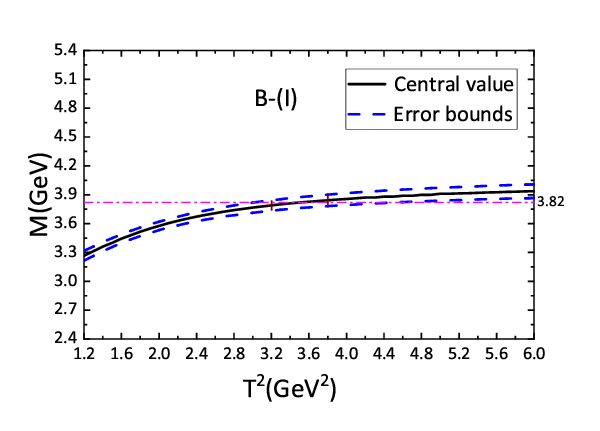}
   \includegraphics[totalheight=5cm,width=7cm]{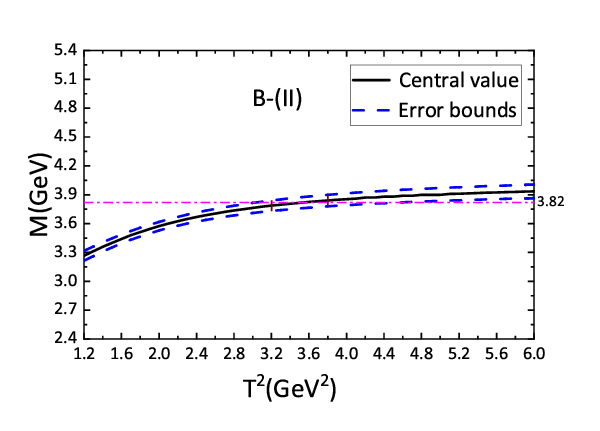}
   \includegraphics[totalheight=5cm,width=7cm]{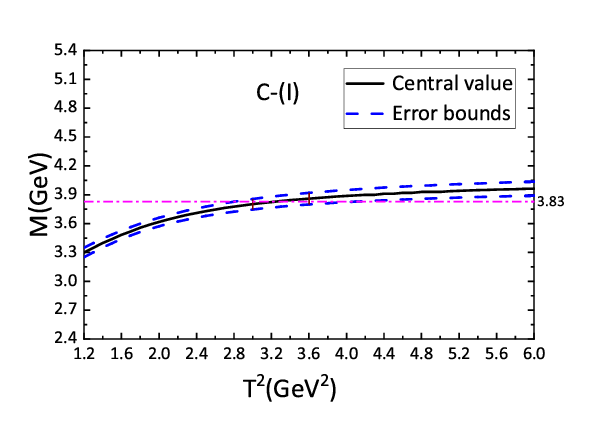}
   \includegraphics[totalheight=5cm,width=7cm]{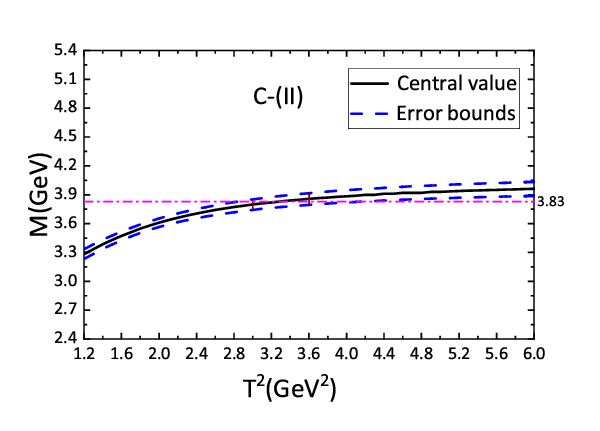}
   \includegraphics[totalheight=5cm,width=7cm]{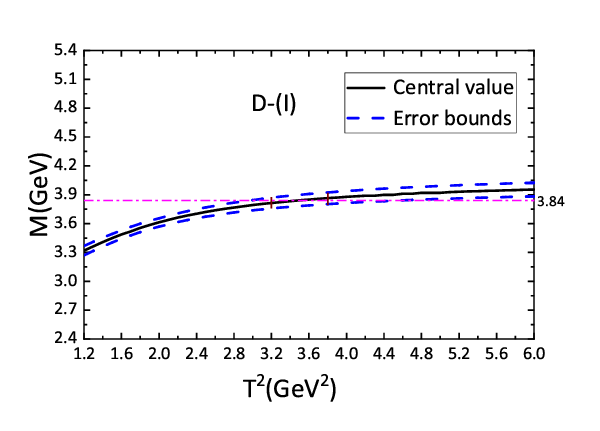}
   \includegraphics[totalheight=5cm,width=7cm]{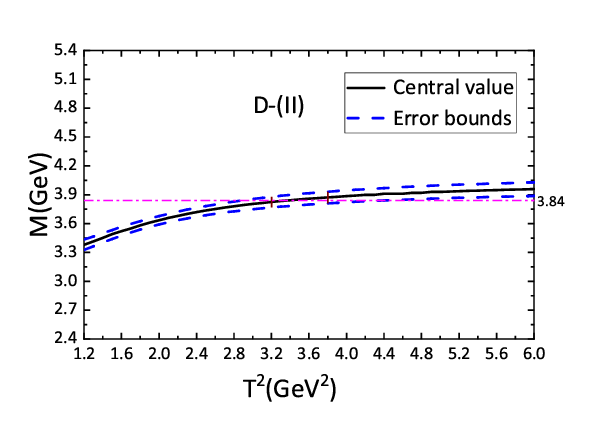}
    \caption{The masses of the charmonium states with variations of the Borel
parameters $T^2$, where the $A$, $B$, $C$ and $D$ denote $\psi_1$, $\psi_2$, $\eta_{c2}$ and $\psi_3$ charmonium states, respectively. }\label{mass-Dwave}
\end{figure}

We plot the charmonium masses with variations  of the Borel parameters $T^2$ in Fig.\ref{mass-Dwave}, the selected ranges of the Borel platforms are shown by the two vertical lines, while the experimental masses  are represented by the horizontal dashed lines. We obtain the central value of the masses based on the pole contribution $50\%$, then consider the uncertainties of the input parameters, and take the pole contribution  $\leq 70\%$ ($\geq 40\%$) to determine the lower (upper) boundary of the Borel windows. It can be seen from Figs.\ref{Dn}-\ref{mass-Dwave} that the contributions  of the vacuum condensates and the masses  remain stable  in the Borel windows.

Although the QCD input parameters I and II differ greatly  in the value of the three-gluon condensate$\langle g_s^3GGG\rangle$,  the  predicted masses are identical. Moreover, the Borel parameter, the continuum threshold parameter and pole are also exactly the same with the only minor deviations  occurring in the decay constants. After taking account  of the  uncertainties, we display the masses and decay constants   in Tables \ref{mass-I}-\ref{mass-II}. The uncertainties in our predictions for the masses are derived by all input parameters through the analytical expressions $\delta =\sqrt{\sum \limits_{i}(\frac{\partial M}{\partial x_i})^2 (x_i-x)^2}$. The uncertainties are propagated using standard error analysis, where $M$ represent the masses, $x_i$ denote the input parameters charm-quark mass, gluon condensates, three-gluon condensates, continuum threshold, and $x$  denote the the central values of these parameters. Since it is difficult to find the analytic expressions of the partial derivatives $\frac{\partial M}{\partial x_i}$, we carry out the approximate numerical calculations and estimate the corresponding uncertainties $\delta \approx \sqrt{\sum \limits_{i}|M(x\pm \Delta x_i) -M(x)|^2}$ via numerical differentiation. To elucidate the origins of the mass uncertainties quoted in Tables \ref{mass-I}-\ref{mass-II}, we examine the individual impacts of the charm-quark mass variation $m_c$, the gluon condensates $\langle\frac{\alpha_s GG}{\pi}\rangle$ and three gluon condensates $\langle g_s^3GGG\rangle$, and the Borel windows $T^2$. Tables \ref{uncertainty-I}-\ref{uncertainty-II} display the resulting masses shift when each parameter is varied independently while others are fixed at their central values.  The gluon condensates contribute negligibly (a few MeV), consistent with the convergent OPE and the perturbative dominance shown in Fig.\ref{Dn} which confirming the reliability of our uncertainties. It is noteworthy that our predicted mass results will be based on the central values  as far as possible and subsequently compared against future experimental or other theoretical results.

\begin{table}
\begin{center}
\begin{tabular}{|c|c|c|c|c|c|c|c|c|}\hline\hline
uncertainty     &$\Delta x_i$      & $\Delta M_{\psi_1(1{}^3D_1)}$    & $\Delta M_{\psi_2(1{}^3D_2)}$    & $\Delta M_{\eta_{c2}(1{}^1D_2)}$    & $\Delta M_{\psi_3(1{}^3D_3)}$
\\ \hline

$m_c$           &$\pm 0.025 \rm{GeV}$                       &$\pm 22.07\rm{MeV}$           &$\pm 23.31\rm{MeV}$       &$\pm 22.58\rm{MeV}$         &$\pm 22.24\rm{MeV}$      \\  \hline

$\sqrt{s_0}$         &$\pm0.1 \rm{GeV}$                   &$\pm 54.54\rm{MeV}$               &$\pm 52.55\rm{MeV}$      &$\pm 53.04\rm{MeV}$      &$\pm 53.75\rm{MeV}$    \\ \hline

$T^2$                &$\pm0.3 \rm{GeV}$                  &$\pm 26.95\rm{MeV}$           &$\pm 29.11\rm{MeV}$       &$\pm 28.85\rm{MeV}$        &$\pm 24.99\rm{MeV}$      \\

\hline\hline
\end{tabular}
\end{center}
\caption{The mass uncertainties  of the parameters variation for input parameter I.}\label{uncertainty-I}
\end{table}

\begin{table}
\begin{center}
\begin{tabular}{|c|c|c|c|c|c|c|c|c|}\hline\hline
uncertainty     &$\Delta x_i$      & $\Delta M_{\psi_1(1{}^3D_1)}$    & $\Delta M_{\psi_2(1{}^3D_2)}$    & $\Delta M_{\eta_{c2}(1{}^1D_2)}$    & $\Delta M_{\psi_3(1{}^3D_3)}$
\\ \hline

$m_c$           &$\pm 0.025 \rm{GeV}$                       &$\pm 21.92\rm{MeV}$           &$\pm 23.51\rm{MeV}$       &$\pm 22.69\rm{MeV}$          &$\pm 22.15\rm{MeV}$       \\  \hline

$\sqrt{s_0}$         &$\pm0.1 \rm{GeV}$                   &$\pm 54.36\rm{MeV}$               &$\pm 52.45\rm{MeV}$          &$\pm 53.06\rm{MeV}$       &$\pm 53.62\rm{MeV}$     \\ \hline

$T^2$                &$\pm0.3 \rm{GeV}$                  &$\pm 26.78\rm{MeV}$           &$\pm 26.67\rm{MeV}$         &$\pm 29.30\rm{MeV}$         &$\pm 23.95\rm{MeV}$       \\ \hline

$\langle\frac{\alpha_s GG}{\pi}\rangle$        &$ \pm0.004 \rm{GeV}^4$
&$\pm 0.22\rm{MeV}$          &$\pm 1.31\rm{MeV}$               &$\pm 0.61\rm{MeV}$          &$\pm 2.61\rm{MeV}$                      \\  \hline

$\langle g_s^3GGG\rangle$       &$\pm0.385 \rm{GeV}^6$
&$\pm 1.45\rm{MeV}$             &$\pm 0.33\rm{MeV}$             &$\pm 1.49\rm{MeV}$             &$\pm 1.68\rm{MeV}$         \\   

\hline\hline
\end{tabular}
\end{center}
\caption{The mass uncertainties  of the parameters variation for input parameter II.}\label{uncertainty-II}
\end{table}

Our prediction $M_{\psi_1}=3.77\pm{0.09}\,\rm {GeV}$ favors assigning  the conventional  $\psi(3770)$ (rather than the $\mathcal{R}(3780)$ reported by the BESIII collaboration \cite{3770-Belle-2004,3770-BESIII-2024}) as the D-wave charmonium $\psi_1$. Our result of $M_{\psi_1(1{}^3D_1)}$  supports the dominance of the D-wave Fock component for $\psi (3770)$. The mass of $\psi (3770)$ lies significantly above $\psi (2S)$. Although the consideration of S-D mixing effects on vector charmonia can interpret their experimental di-electric widths \cite{LYR-3770-2024}, the S-wave admixture were very large, such agreement would be unlikely. The predicted mass $M_{\psi_2}=3.82\pm{0.09}\,\rm {GeV}$  is consistent with the observation of the Belle, BESIII and LHCb collaborations \cite{Belle-3823-2013,BESIII-3823-2015,LHCb-3823-2020,BESIII-3823-2022-1,
BESIII-3823-2022-2}, and favors the identification $\psi_2$ indeed. While the $1{}^1D_2$ charmonium state has not been discovered yet, we estimate the mass to be  $M_{\eta_{c2}}=3.83\pm{0.09}\,\rm {GeV}$, which is expected  to be observed at the BESIII, LHCb, Belle II collaborations in the future. Furthermore, we obtain the mass $M_{\psi_3}=3.84\pm{0.08}\,\rm {GeV}$, and tentatively identify the $X(3842)(\psi_3(3842))$  as the $1{}^3D_3$ charmonium state \cite{3842-LHCb-2019}.
We can take the decay constants $f_{\psi_1/\psi_2/\eta_{c2}/\psi_3}$ are elementary input parameters to study the strong decays and radiative decays to make the identifications in a more robust way.

The comparison of our mass predictions with those from other theoretical approaches, as summarized in Table \ref{mass-D-wave-charmonia}. The potential models generally predict masses in the range of $3.77-3.84$~GeV, if we consider the uncertainty ranges, the predictions from these methods  fall within our calculated error bands. Therefore, the following discussion will focus on comparison of the central values, for the $\psi_1$ state, the quark model predictions are generally higher than our results, which may be attributed to inherent model dependencies. Our central values of the unobserved $\eta_{c2}$ is closed to some quark models. This concordance between the methods is particularly noteworthy and reinforces the credibility of approaches. Further experimental data are essential to advance our understanding of the D-wave charmonium systems. In particular, the observation of the $\eta_{c2}$ state and more precise measurements of masses, widths, and decay channels for the charmonia are crucial to test theoretical predictions.

\begin{table}
\begin{center}
\begin{tabular}{|c|c|c|c|c|c|c|c|c|}\hline\hline
\multirow{2}{*}{}   &\multirow{2}{*}{ref.}       & $M_{\psi_1(1{}^3D_1)}$            & $M_{\psi_2(1{}^3D_2)}$    & $M_{\eta_{c2}(1{}^1D_2)}$   & $M_{\psi_3(1{}^3D_3)}$  \\
&   &(\rm{GeV})       &(\rm{GeV})    &(\rm{GeV})  &(\rm{GeV})  \\  \hline
\multirow{1}{*}{ QCD sum  rules }   &this work        &$3.77$      &$3.82$   &$3.83$  &$3.84$        \\ \hline
\multirow{1}{*}{Bethe-Salpeter equations}
&\cite{Fischer-charmonia-2015}        &-           &$3.739$      &$3.806$     &$3.896$  \\
\hline
\multirow{2}{*}{nonrelativistic potential model}
&\cite{Barnes-D-wave-charmonia-2005}  &$3.785$    &$3.800$      &$3.799$     &$3.806$  \\
&\cite{CKT-D-wave-charmonia-2009}     &$3.787$    &$3.798$      &$3.796$     &$3.799$  \\

\multirow{1}{*}{(linear potential model)}
&\cite{ZXH-D-wave-charmonia-2017}     &$3.787$    &$3.807$      &$3.806$     &$3.811$  \\
\multirow{1}{*}{(screened potential model)}
&\cite{ZXH-D-wave-charmonia-2017}     &$3.792$    &$3.807$      &$3.805$     &$3.808$  \\
\hline
\multirow{1}{*}{unquenched potential model }
&\cite{LX-D-wave-charmonia-2019}      &$3.830$    &$3.848$      &$3.848$     &$3.859$  \\
\hline
\multirow{1}{*}{GI model }
&\cite{Barnes-D-wave-charmonia-2005}  &$3.819$    &$3.838$      &$3.837$     &$3.849$  \\
\hline
Coulomb plus &\multirow{2}{*}{\cite{Chaturvedi-D-wave-charmonia-2018}}     &\multirow{2}{*}{$3.815$}    &\multirow{2}{*}{$3.814$}      &\multirow{2}{*}{$3.806$}     &\multirow{2}{*}{$3.798$}  \\
linear potential model  &&&&&
 \\ \hline
\multirow{1}{*}{Cornell potential model}
&\cite{Soni-D-wave-charmonia-2018}    &$3.775$     &$3.772$      &$3.765$     &$3.755$ \\ \hline
\multirow{1}{*}{Cornell coupled-channel model }
&\cite{Eichten-2003}                  &-           &$3.831$      &$3.838$     &$3.868$ \\ \hline
Cornell potential coupled &\multirow{2}{*}{\cite{Chaturvedi-D-wave-charmonia-2020}}     &\multirow{2}{*}{$3.785$}    &\multirow{2}{*}{$3.800$}      &\multirow{2}{*}{$3.780$}     &\multirow{2}{*}{$3.806$}  \\
relativistic correction &&&&&
 \\ \hline
QCD-motivated relativistic  &\multirow{2}{*}{\cite{Ebert-D-wave-charmonia-2011}}     &\multirow{2}{*}{$3.783$}    &\multirow{2}{*}{$3.795$}      &\multirow{2}{*}{$3.807$}     &\multirow{2}{*}{$3.813$}  \\
quark model &&&&&
 \\  \hline
\multirow{2}{*}{other potential model}
&\cite{Sultan-D-wave-charmonia-2014}   &$3.781$      &$3.800$      &$3.799$     &$3.805$ \\
&\cite{Radford-D-wave-charmonia-2007}  &$3.804$      &$3.824$      &$3.824$     &$3.831$ \\
\hline
\hline
\end{tabular}
\end{center}
\caption{The  mass of the D-wave charmonia.}\label{mass-D-wave-charmonia}
\end{table}

The decay constant is a crucial parameter in the calculation of strong, leptonic, semileptonic, and radiative decays. For instance, in Ref.\cite{Lujie-semi-charm-2026}, the decay constants $f_{\psi_{1}}$, $f_{\psi_{2}}$, $f_{\eta_{c2}}$, $f_{\psi_{3}}$  presented in this work is employed within the framework of three-point correlation functions to compute the semileptonic decay width of D-wave charmonia. There is a scarcity of research dedicated to direct computate the decay constants for D-wave charmonia \cite{Krassnigg-decayconstans-charm-2018}. Our calculated constants are not merely numerical outputs but are key phenomenological parameters that bridge the gap between mass predictions and the future analysis of decay widths and branching ratios.

The values of the non-perturbative vacuum condensates, particularly the triple-gluon condensate $\langle g_s^3GGG\rangle$, are not as precisely determined as the fundamental parameters like the charm quark mass $m_c$. Different extraction methods from QCD sum rules analyse or phenomenological fits can lead to differing numerical estimates. The predicted masses for all four D-wave charmonium states remain remarkably stable and consistent between the two parameter sets, as detailed in Tables 3 and 4, the variations are well within the quoted uncertainties. This demonstrates that our mass predictions are dominated by the perturbative contribution and are not sensitive to the specific choice of the gluon condensate values and the triple-gluon condensate. While the masses are robust, we observe the minor sensitivity in the decay constants, the parameter sets I and II are considered.

\section{Conclusion}
In this work, we perform a systematic analysis of the conventional D-wave charmonium states $\psi_1$, $\psi_2$, $\eta_{c2}$ and $\psi_3$ in the framework of the QCD sum rules. We carry out the operator product up to the condensates of dimension-6, which includes both the gluon condensate $\langle \frac{\alpha_s GG}{\pi} \rangle$ and the three-gluon condensate $\langle g_s^3 GGG \rangle$. Notably, despite significant differences between the values of the three-gluon condensate in the two adopted parameter sets I and II  adopted, the resulting mass predictions remain remarkably stable and consistent. This robustness underscores the reliability of the QCD sum rules and the dominant role of the perturbative contributions. We observe that the predicted  masses favor identifying the $\psi(3770)$, $\psi_2(3823)$ and $\psi_3(3842)$ as the D-wave charmonium states $1{}^3D_1$, $1{}^3D_2$ and $1{}^3D_3$, respectively. Since the $\eta_{c2}(1{}^1D_2)$ has not been observed experimentally yet, we make prediction to be confronted to experimental data  in the future. Meanwhile,the obtained decay constants can serve as elementary input parameters in subsequent study of the strong decays, leptonic and radiative decays to make the identifications in a more robust way.

\section*{Acknowledgements}
This  work is supported by National Natural Science Foundation, Grant Number 12175068 and the Fundamental Research Funds for the Central Universities (2025MS177).

\end{document}